\documentclass[nofootinbib,aps]{revtex4}

\usepackage{amstext,amsmath,amssymb}
\usepackage[dvips]{graphicx}
\usepackage{latexsym}
\usepackage{epsfig}

\def \v{\vspace}

\newcommand{\N}{\mathbb{N}}
\newcommand{\Z}{\mathbb{Z}}

\newcommand{\R}{\mathbb{R}}
\newcommand{\C}{\mathbb{C}}

\def\be{\begin{equation}}
\def\ee{\end{equation}}
\def\bes{\begin{eqnarray}}
\def\ees{\end{eqnarray}}
\def\nn{\nonumber}
\def\arr{\rightarrow}

\def\la{\langle}
\def\ra{\rangle}
\def\f{\frac}

\def\what{\widehat}

\def \vphi{\varphi}

\def\hh{{\cal H}}
\def\mm{{\cal M}}
\def\ss{{\cal S}}
\def\aa{{\cal A}}

\def\ct{\cos\theta}
\def\st{\sin\theta}

\newcommand{\lalg}[1]{\mathfrak{#1}}
\newcommand{\SU}{\mathrm{SU}}
\newcommand{\SL}{\mathrm{SL}}
\newcommand{\SO}{\mathrm{SO}}
\newcommand{\U}{\mathrm{U}}
\newcommand{\su}{\lalg{su}}

\newcommand{\so}{\lalg{so}}

\renewcommand{\v}{\overrightarrow}

\newcommand{\ket}[1]{|#1\rangle}

\newcommand{\bk}[1]{\langle #1 \rangle}

\begin{document}

\title{Coherent States for 3d Deformed Special Relativity: \\
semi-classical points in a quantum flat spacetime}

\author{{\bf Etera R. Livine}\footnote{elivine@perimeterinstitute.ca}}
\affiliation{Perimeter Institute, 35 King Street North
Waterloo, Ontario, Canada N2J 2W9}
\author{{\bf Daniele Oriti}\footnote{d.oriti@damtp.cam.ac.uk}}
\affiliation{Department of Applied Mathematics and Theoretical Physics,
Centre for Mathematical Sciences, University of Cambridge,
Wilberforce Road, Cambridge CB3 0WA, UK,}
\affiliation{Girton College, University of Cambridge, Cambridge CB3 0JG, UK}

\begin{abstract}

\begin{center}
{\small ABSTRACT}
\end{center}
We analyse the quantum geometry of 3-dimensional deformed special relativity (DSR) and the notion of spacetime points in such a context, identified with coherent states that minimize the uncertainty relations among spacetime coordinates operators. We construct this system of coherent states in both the Riemannian and Lorentzian case, and study their properties and their geometric interpretation.      

\end{abstract}

\maketitle

\tableofcontents

\section{Introduction}

Motivated mainly by the possibility of Quantum Gravity phenomenology
\cite{QGphen}, Deformed (or Doubly) Special Relativity \cite{Giovanni}
has emerged recently as a possible description of the kinematics of
Quantum Gravity, in the flat space approximation. The precise relation
with the full theory, still to be constructed, is not yet fully
understood \cite{jurekQG}; however, in the 3d case solid arguments have been obtained for this to be the
case \cite{jll,agl}, and even a rigorous derivation of some characterizing DSR features from a speicifi quantum gravity model has been obtained \cite{sf3d}, and in the 4d case, although the situation is much less clear, interesting arguments have been found
\cite{agl,usflorian}, confirming the reasons for the current interest in this model.

The basic motivation for constructing such a model is the
attempt to incorporate the presence of a non-zero invariant minimal length
(thought of as a remnant at the kinematical level of the underlying
quantum gravity theory, and therefore usually identified with the Planck
length) into the kinematical setting of Special Relativity, where the
other invariant quantity, the speed of light, is present.
The outcome of this deformation of Special Relativity to accommodate
two invariant scales is first of all a modification of the symmetry
group of the theory, that becomes the $\kappa$-Poincar\'e group \cite{lukierski}, and a consequent modification of dispersion relations for
moving particles, of reaction thresholds of scattering processes, and a
vast arena of new possibilities for testing quantum gravity models
\cite{DSRphen}.

A deformed symmetry group for spacetime can be equivalently
seen \cite{Majid} as a non trivial metric on momentum space, that
becomes curved and, more precisely, in the DSR case endowed with a De
Sitter (or Anti-DeSitter) geometry. This has important consequences
for the geometry of spacetime. Defining spacetime \lq\lq coordinates'' $X_\mu$
as Lie derivatives over this curved momentum space, in other words
\lq\lq by duality'' with respect to momenta, they turn out to be
non-commutative, being defined (at least in one particular choice of
basis \cite{snyder}) by the Cartan decomposition of the
de Sitter algebra into its Lorentz subalgebra and the remnant (the
spacetime coordinates themselves). Also, they have a non-trivial action on
momentum space, characterized by a deformation parameter
$\kappa=1/l_p$, where $l_p$ is the Planck length, i.e. the same
deformation parameter appearing in the $\kappa$-Poincar\'e
algebra. Spacetime acquires then, also in this approximately flat
effective description, a quantum nature, and its geometry must be
dealt with accordingly. Indeed, the quantities $X_\mu$ are quantum {\it operators}, first
of all, and their representations must be studied in order to study
the properties of spacetime in this model, and are best understood not
as \lq\lq coordinates'' but as basic distance operators in spacetime
(with respect to a given origin), due also to their transformation
properties under Lorentz transformations (they transform as
4-vectors). The study of their properties (eigenstates, spectra, etc),
and more generally of the
quantum spacetime geometry of DSR, has been started in
\cite{discrete}, using tools from the representation theory of the
Lorentz group in various dimensions.
Another issue that was discussed in \cite{discrete} is the notion of
\lq\lq spacetime point'' in a quantum spacetime, with its quantum nature
reflected in the non-commutativity of the distance operators $X_\mu$
that we would use to localize points. The best definition available of
a spacetime point in this quantum setting is that of a coherent state
in the Hilbert space of quantum states of geometry, being the
representation space of the distance operators $X_\mu$,
which minimizes the uncertainty in all the $X_\mu$ themselves. 
Indeed, we take this as a definition of a 'point' in a quantum spacetime.

In general
one would expect a minimal simultaneous uncertainty, i.e. an uncertainty in the measurement of several spacetime coordinate, as for measuring a spacetime distance $l$, of the form:
\be
\delta l \ge l_P \times \left(\f{l}{l_P}\right)^{\alpha}
\label{delta}
\ee
with $\alpha=1/2$ in 3d and $\alpha=1/3$ in 4d assuming the validity of the holographic principle, or $\alpha=1/3$ in 3d and $\alpha=1/4$ in 4d, not making this assumption (the argument for this was reviewed in \cite{discrete}).

The exact form of the uncertainty relations is basis dependent in DSR, i.e. it depends on the
precise definition of the $X_\mu$ operators \cite{jurek} one chooses.
In the so-called $\kappa$-Minkowski basis, in which the $X_\mu$
form a Lie algebra, the notion of coherent state and minimal
uncertainty is straightforward to implement and was indeed obtained in
\cite{discrete}; however, in this basis the interpretation of the
$X_\mu$ as distance operators is not immediate and the geometric
meaning of the is more obscure. On the other hand, in the Snyder
basis, to which we were referring above, obtained by Cartan
decomposition, this interpretation is clear, but the construction of
coherent states, i.e. the identification of what is a spacetime point,
is more technically involved and the framework of generalized coherent
states on Lie groups \cite{cohstates} has to be used.

The procedure for constructing a system of coherent states is (briefly) the following: given a Lie Group $G$ and a representation $T$ of it acting on the Hilbert space $\cal{H}$, with a fixed vector $\mid \Psi_0\rangle$ in it, the system of states $\left\{ \mid \Psi_g\rangle = T(g)\mid \Psi_0\rangle\right\}$ is a system of coherent states.
The point is then to identify which are the states  $\Psi_0$ that are closest to classical, in the sense that they minimize the invariant dispersion or variance $\Delta = \Delta C = \langle \Psi_0 \mid C \mid \Psi_0 \rangle - g^{ij} \langle \Psi_0 \mid X_i \mid \Psi_0 \rangle \langle \Psi_0 \mid X_j \mid \Psi_0 \rangle$, where $C=g^{ij}X_i X_j$ is the invariant quadratic Casimir of the group $G$.
The general requirement is the following: call $\lalg{g}$ the algebra of $G$, $\lalg{g}^c$ its complexification, $\lalg{b}$ the isotropy subalgebra of the state $\mid \Psi_0\rangle$, i.e. the set of elements $b$ in $\lalg{g}^c$ such that $T_b \mid \Psi_0\rangle = \lambda_b \mid \Psi_0\rangle$, with $\lambda_b$ a complex number, and $\bar{\lalg{b}}$ the subalgebra of $\lalg{g}^c$ conjugate to $\lalg{b}$; then the state $\mid \Psi_0\rangle$ is closest to the classical states if it is most symmetrical, that  is if $\lalg{b}\oplus\bar{\lalg{b}} = \lalg{g}^c$, i.e. if the isotropy subalgebra $\lalg{b}$ is maximal.

In this paper, we tackle exactly the problem of defining points in a quantum spacetime with the formalism of coherent states using these general
techniques. In the first part of the paper we construct coherent states minimizing the spacetime uncertainty relations (thus representing semi-classical points) for 3d DSR, first in the Euclidean setting and then in the Minkowskian one, our construction being based on the fact that the momentum space in 3d DSR is given by the $\SU(2)$ group manifold in the Euclidean and by the $\SU(1,1)$ group manifold in the Minkowskian; then we show a similar construction based instead on the homogeneous space $\SO(4)/\SO(3)$ in the Euclidean and on $\SO(3,1)/\SO(2,1)$ in the Minkowskian, that gives similar results and represents in a slightly simplified context the same construction that one would perform for 4d DSR. We confine ourselves to the 3-dimensional case purely for greater simplicity of the relevant calculations, which become much more
involved in higher dimensions; however we stress that no major conceptual or
technical difference is expected in the more realistic case of 4
spacetime dimensions, or in higher ones for what matters; these cases
can be in principle handled using the same tools and with the same
procedure.

\section{Coherent States: The Euclidean 3d space}

In 3d deformed special relativity (DSR), the momentum space $\mm$ is simply the group manifold $\SU(2)$. The Hilbert space of wave functions (say, for a point particle living in the DSR spacetime) in the momentum polarization is then the space of $L^2$ functions over $\SU(2)$ equipped with the Haar measure:
$\hh\equiv L^2(\SU(2),dg)$. In the framework of spin foam quantization, it has been shown that this curved momentum space structure can be derived directly from 3d (Riemannian) quantum gravity coupled to matter \cite{sf3d}. More precisely, the effective theory describing the matter propagation, after integration over the gravity degrees of freedom, is a DSR theory invariant under a $\kappa$-deformation of the Poincar\'e algebra, where the deformation parameter $\kappa$ is simply related to the Newton constant for gravity and therefore to the Planck length.

In this context, the curved momentum space naturally leads to a deformation of the rule of addition of momenta. Indeed the addition of momenta will not be the simple $\R^3$ addition anymore, but will be defined by the multiplication on the $\SU(2)$ group. More technically, the standard convention is to define the 3-momentum $\v{p}$ from the $\SU(2)$ group element as follow:
\be
g(\v{p})\equiv
\sqrt{1-\f{\v{p}^2}{\kappa^2}} \,{\rm Id}\,
+i \f{\v{p}}{\kappa}.\,\v{\sigma},
\ee
where $\v{\sigma}$ are the usual Pauli matrices and $\kappa$ the Planck mass. Then the deformed addition of momenta $\oplus$ is given by:
\be
g(\v{p}_1\oplus \v{p}_2)\,\equiv\, g(\v{p}_1)g(\v{p}_2),
\ee
or more explicitly:
$$
\v{p}_1\oplus \v{p}_2=
\v{p}_1\sqrt{1-\f{\v{p}_2^2}{\kappa^2}}
+\v{p}_2\sqrt{1-\f{\v{p}_1^2}{\kappa^2}}
-\f{1}{\kappa}\v{p}_1\wedge \v{p}_2.
$$
This leads to a deformation of the law of conservation of energy-momentum, which is interpreted as due to the gravitational interaction.

\medskip

In the present work, we are interested by the dual structure to the momentum space, and we would like to describe the space(-time) structure of the 3d DSR theory. Given the $SU(2)$ group structure of momentum space, there is a preferred choice for the coordinates, that in a sense combines the virtues of the Snyder basis and of the $\kappa$-Minkowski basis for 4d DSR. The coordinates $X_i$ are now operators acting on the Hilbert space $\hh$ and more precisely, they are the translation operators on the group manifold $\SU(2)$, as in the Snyder basis: $X_i\equiv l_P J_i$ are the standard $\su(2)$ algebra generators, which are represented by the Pauli matrices $\v{\sigma}$ in the fundamental two-dimensional representation, and $l_P\equiv \hbar/\kappa$ is the 3d Planck length.

The spacetime coordinates are now non-commutative:
\be
[X_i,X_j]\,=\, il_P\,\epsilon_{ijk}X_k,
\ee
and obviously form a Lie algebra, just as in the $\kappa$-Minkowski basis of 4d DSR,
and also the bracket between positions and momenta is deformed:
\be
[X_i,p_j]\,=\, -i\hbar\sqrt{1-\f{\v{p}^2}{\kappa^2}}\,\delta_{ij}+ il_P\,\epsilon_{ijk}p_k.
\ee

The (squared) spacetime length operator $L^2\equiv X_iX_i$ is actually the Casimir operator of $\SU(2)$. It is invariant under $\SU(2)$ and commutes with the coordinates $X_i$. Its spectrum is of course $l_P^2\times j(j+1)$, with $j\in\Z$, while the spectrum of the $X_i$ coordinate operators is simply $l_P\,\Z$. This way, one can view this non-commutative spacetime as implementing a minimal length $l_P$ and a discrete structure of spacetime. Note also that this quantum spacetime structure is analogous to the one discovered in 3d loop quantum gravity and spin foam models, and was indeed studied as a simplified version of it, with its Lorentz invariance analysed in detail, in \cite{discrete}. 

Because the spacetime coordinates do not commute with each other anymore, we can not simultaneously diagonalize the
$X_i$'s and localize the three coordinates of a point/particle perfectly at the same time. So we face the issue of how to define a point. In order to be able to talk about a spacetime point, as it was proposed in \cite{discrete}, we need to introduce coherent states minimizing the uncertainty in the $X_i$'s, which will define a notion of semi-classical points in a non-commutative geometry. Actually, coherent states for Lie groups, and especially $\SU(2)$, are well-known  and we will review the material presented by Perelomov \cite{cohstates} from our own perspective; coherent states for the algebra generated by the $X_i$ operators, named \lq\lq the fuzzy sphere\rq\rq and considered as a paradigmatic example of a non-commutative spacetime, were constructed using a different approach in \cite{majidCoh}.

\subsection{A first length uncertainty estimate}

Starting with the $\su(2)$ algebra $[X_i,X_j]\,=\, il_P\,\epsilon_{ijk}X_k$, we get the following uncertainty relations for $i\ne j$:
\be
(\delta X_i)^2(\delta X_j)^2\ge \f{l_P^2}{4}\la\epsilon_{ijk}X_k\ra^2.
\label{uncertaintyij}
\ee
Summing all these uncertainty relations over $i$ and $j$ (possibly equal), we obtain\footnote{We would like to thank Laurent Freidel for pointing out this simple fact to us.}:
$$
(\delta l)^4\ge 2\f{l_P^2}{4}\,l^2 + \sum_i (\delta X_i)^4\ge \f{l_P^2}{4}\,2l^2,
$$
where we have introduced the distance $l^2\equiv \sum_i \la X_i\ra^2$ and the uncertainty $(\delta l)^2\equiv \sum_i (\delta X_i)^2$. Finally, we have derived the following uncertainty relation:
\be
(\delta l)^2\ge \f{1}{\sqrt{2}}l_Pl.
\ee
This provides us with a first bound on the uncertainty, which suggests a generic feature of the localization of spacetime points: we can not localize a space point at a distance $L$ from us with an uncertainty smaller than of the order of $\sqrt{l_PL}$.
Actually, computing the uncertainty exactly over the quantum states, we will find in the following that the true minimal spread is simply $(\delta L)^2\ge l_P L$. This is simply due to the term $\sum_i (\delta X_i)^4>0$.

\subsection{$\SU(2)$ coherent states and the minimal uncertainty}
States diagonalizing $Z\equiv X_3$ are the usual $|j,m\ra$, where $j$ is the spin label and $m$ the angular momentum projection over the $X_3$ direction. It is easy to check the expectation values of the coordinate operators on such states:
\be
\la X_1\ra_{j,m}=\la X_2\ra_{j,m}=0,
\qquad
\la X_3\ra_{j,m}=m\,l_P.
\ee
Therefore such states correspond geometrically to points located along the $X_3$ axis at (non-invariant) distance $l_P m$.

The spread of the state or its uncertainty (which determines the precision of its localization) is defined as:
\be
\delta L = \sqrt{|\la X_iX_i\ra-\la X_i\ra\la X_i\ra|}.
\ee
Then we easily compute:
\be
(\delta L)_{j,m}=\,l_P\,\sqrt{j(j+1)-m^2},
\ee
which is minimal for $m=j$ (and for $m=-j$). In other words, the maximal (or minimal) weight states $|j,j\ra$ (or $|j,-j\ra$) are the coherent states for $\SU(2)$ minimizing the uncertainties in the positions and thus are the semi-classical states for 3d Euclidean DSR, i.e. the best definition of spacetime points in the corresponding quantum spacetime. Of course, these correspond to \lq\lq some\rq\rq of the points of this quantum spacetime, those at the location mentioned above. On such states, we have:
$$
\la L\ra=\sqrt{\la X_i X_i\ra}=l_P\sqrt{j(j+1)},
\quad \textrm{and}\quad
\delta L\,=\,l_P\sqrt{j}=\,l_P\,\left(\sqrt{\f{L^2}{l_P^2}+\f{1}{4}}-\f{1}{2}\right),
$$
which confirms the fact that these quantum points are approximately localized at a spacetime distance $l_p j$ from the origin, i.e from the location of the observer, and we derive the relation for large distance\footnotemark:
\be
\delta L\,\sim\,\sqrt{L l_P}.
\ee
\footnotetext{We could have defined the average distance from the origin through Pythagoras formula  as $\la L\ra\equiv \sqrt{\la X_i\ra\la X_i\ra}$ which would lead to $\la L\ra=\,j\,l_P$. In this case, the relation between the distance and the uncertainty simplifies to:
$$
\delta L\,=\,l_P\sqrt{j}=\sqrt{L l_P},
$$
and we recover exactly the bound set by the holographic principle.}

This relation is exactly the one expected from the holographic principle in 3d, as mentioned in the introduction. We stress that the uncertainty is actually substantially larger than the Planck length $l_P$, as expected  from several arguments about quantum measurements in the context of general relativity \cite{ng2}.

Now we want to define all the other points of our quantum spacetime, i.e. construct all the other coherent states minimizing the spacetime uncertainty relations, following the Perelomov construction, according to which the full coherent states system is obtained from the vector invariant under the maximal subalgebra by the action of a group transformation.    
More technically, we define the (coherent) states:
\be
|j,\what{n}\in\ss^2\ra \equiv
e^{\alpha J_+ -\bar{\alpha}J_-}|j,j\ra
=\sum_{m=-j}^{+j}
\left(\f{(2j)!}{(j+m)!(j-m)!}\right)^{\f{1}{2}}\f{z^{j-m}}{(1+|z|^2)^j}\,|j,m\ra,
\ee
with the following notations:
$$
\what{n}=(\sin\theta\cos\phi,\sin\theta\sin\phi,\cos\theta),
\quad \alpha=\f{\theta}{2}e^{-i\phi},
\quad z=\tan\f{\theta}{2}e^{-i\phi}.
$$
Using the explicit expression, one can check that:
$$
\la X_1 \ra_{j,\what{n}}=2j\f{{\rm Re}(z)}{1+|z|^2}, \quad
\la X_2 \ra_{j,\what{n}}=2j\f{{\rm Im}(z)}{1+|z|^2}, \quad
\la X_3 \ra_{j,\what{n}}=j\f{1-|z|^2}{1+|z|^2},
$$
or more simply:
\be
\la \v{X} \ra_{j,\what{n}}\,=\,j\,\what{n} \,l_P.
\ee
This means that the above states corresponds to the points located at a distance $l_P j$ in the direction $\what{n}$ from the origin, and the set of all these states spans the 2-sphere of radius $l_P j$ centered in the origin.
We can compute the overlap between two coherent states and we obtain:
\be
\la j,\what{n_1}|j,\what{n_2}\ra=
e^{i\Phi(\what{n_1},\what{n_2})}\left(\f{1+\what{n_1}\cdot\what{n_2}}{2}\right)^j,
\ee
where the phase is
$$
\Phi(\what{n_1},\what{n_2})=j\,\aa(\what{n_0},\what{n_1},\what{n_2})
\,=\,\f{j}{2i}\ln\left(\f{1+\alpha_1\overline{\alpha_2}}{1+\overline{\alpha_1}\alpha_2}\right),
$$
with $\aa(\what{n_0},\what{n_1},\what{n_2})$ the area of the geodesic triangle on the sphere $\ss^2$ with vertices given by $\what{n_0}=(0,0,1)$, $\what{n_1}$ and $\what{n_2}$.

These states are truly the $\SU(2)$ semiclassical coherent states, meaning that they saturate (minimize) the uncertainty relation
$$
(\delta X_1)(\delta X_2)\ge \f{l_P}{2}\la X_3\ra,
$$
induced by the commutation relation $[X_1,X_2]=il_P\,X_3$, as well as the others obtained by cyclic permutations of these coordinate operators. And finally, they form a resolution of unity\footnotemark:
\be
\f{2j+1}{4\pi}\,\int {\rm d}^2\what{n}\,|j,\what{n}\ra\la j,\what{n}|\,=\,
{\rm Id}_j,
\ee
\footnotetext{The measure on the sphere is given by:
$$
{\rm d}^2\what{n}\equiv \sin\theta {\rm d}\theta {\rm d}\phi
=\f{4{\rm d}^2z}{(1+|z|^2)^2}.
$$}
where ${\rm Id}_j$ is the identity on the Hilbert space $\hh_j$ of the spin-$j$ representation (whose basis vectors are the $|j,m\ra$, $-j\le m\le +j$). This confirms (it is the algebraic characterization of this) that they exhaust all the points of the 2-sphere of radius $l_p j$ around the origin of our quantum spacetime. From this one can prove that the set of {\it all} states of this form, for {\it all} values of $j$, span the whole quantum spacetime under consideration; in other words, one can prove, using the Peter-Weyl measure for weighting each representation $j$, that this system of coherent states provides us with a resolution of unity on the full Hilbert space $\hh$:
$$
{\rm Id}=\f{1}{4\pi}\sum_j
(2j+1)\,\int {\rm d}^2\what{n}\,|j,\what{n}\ra\la j,\what{n}|.
$$
This is actually a positive-operator-valued-measure (POVM). Indeed the projection operator $|j,\what{n}\ra\la j,\what{n}|$ allows the simultaneous measurement of the three spacetime coordinates $X_1,X_2,X_3$, even though this measurement is not sharp since the states $|j,\what{n}\ra$ have a non-vanishing overlap. These operators allow us to localize a spacetime point with an uncertainty $\delta L \sim \sqrt{L}$ depending on the distance $L$ to the origin. This gives a special role to the observer, who is located at the origin. We also note that a similar analysis of the (quantum) point structure of a non-commutative spacetime in terms of POVMs was performed by Toller in \cite{toller}. 

We would like to end this section on the remark that to go from one 'point' $|j,\what{n_1}\ra$ of our non-commutative spacetime to another point $|j,\what{n_2}\ra$, we actually simply do a $\SU(2)$ rotation $g$, which takes $\what{n_1}$ to $\what{n_2}$ i.e. we use the position operators $X_i=l_P\,J_i$ themselves to generate a spacetime translation. This let us move along the 2-sphere of of radius $l_P j$. Nevertheless, we see that these special translations do not allow us to change the value of the distance $j$ and to go from one state at $j_1$ to another at $j_2$: for this we would need the momentum operators $p_i$. We discuss briefly below why the action of the momentum operators on these states is not easy to analyze.

\medskip

Let us sum up the situation for this Euclidean 3d deformed special relativity. The momentum space is $\SU(2)$ and the Hilbert space $\hh$ of wave functions of the theory is $L^2(\SU(2)$. An orthogonal basis of $\hh$ is provided by the matrix elements of the group elements in all the representations of $\SU(2)$, we note them
$D^j_{ab}(g)\equiv \la j,a|g|j,b\ra$:
$$
\la D^k_{cd}|D^j_{ab}\ra\equiv
\int dg D^j_{ab}(g)\overline{D^k_{cd}(g)}
=\f{\delta_{jk}}{2j+1}\delta_{ac}\delta_{bd}.
$$
To normalize this states, we multiply them by a factor $\sqrt{2j+1}$.
Functions of the momentum $g$ (or $\v{p}$) acts by multiplication
$$
\what{\vphi(\v{p})}.D^j_{ab}(g) \equiv \vphi(\v{p}(g))D^j_{ab}(g),
$$
while the position operators $\vec{X}$ acts as derivations
$$
\what{X_i}.D^j_{ab}(g) \equiv D^j_{ab}(l_P J_i\cdot g)=l_P\,\la j,a|J_ig|j,b\ra,
$$
or equivalently
$$
\what{e^{i\v{v}\cdot\v{X}}}.D^j_{ab}(g) \equiv D^j_{ab}(e^{il_P\,\v{v}\cdot\v{J}}g).
$$
As the operator $\what{X_i}$ acts on the left, it acts only on $a$ and leaves $b$ invariant. Therefore, keeping $b$ fixed, the whole previous analysis of $\SU(2)$ coherent states applies to the vector $|j,a\ra$: states having a minimal spacetime spread are $|D^j_{\what{n}b}\ra\,\equiv\, \la j,\what{n}|g|j,b\ra$. Now, an important remark is that the standard translations $e^{i\v{v}.\v{p}}$ do not map a coherent state to another coherent state, or that the system of coherent states is not invariant under the action of the standard translations. In a sense we can thus say that the system of coherent states $\{|j,\what{n}\ra, j\in\N/2, \what{n}\in\ss^2\}$ is relative to a given observer (located at the origin). More precisely, $e^{i\v{v}\cdot\v{p}}$ acts simply by multiplication on the matrix elements:
\be
\what{e^{i\vec{v}\cdot\vec{p}}}\cdot D^j_{ab}(g)
=e^{i\vec{v}\cdot\vec{p}(g)}\,D^j_{ab}(g)
=e^{\f{i\kappa}{2}\vec{v}\cdot{\rm Tr}(g\vec{\sigma})}\,D^j_{ab}(g),
\ee
where we take the trace in the fundamental two-dimensional representation, but the function $\exp(i\vec{v}\cdot{\rm Tr}(g\vec{\sigma}))$ decomposes into matrix elements of all possible representations of $\SU(2)$. On the other hand, we can introduce the simpler operators given by multiplication by the matrix elements themselves $D^k_{cd}(g)$. However we can not interpret them as translation operators in a straightforward way. Overall, it appears that it is not straightforward at all to define translation operators that have a simple action on the coherent states.

\subsection{About plane waves in non-commutative geometry}

Coherent states for this Euclidean spacetime based on the $\su(2)$ algebra representing semi-classical points were actually already proposed in \cite{majidCoh} (although the authors did not recognize them as the highest weight vectors $g\,|j,j\ra$ as written by Perelomov), where the authors also investigate the possibility of writing a Poincar\'e covariant differential calculus. A topic of special interest is the plane waves. More precisely, from the perspective of measuring the difference between a plane wave in flat commutative spacetime from a plane wave in non-commutative space, it is interesting to compare the expectation value $\la \exp(i\v{p}.\v{X})\ra$ for fixed momentum $\v{p}$ on a coherent state to its classical value $\exp(i\v{p}.\la\v{X}\ra)$.

Explicitly computing these expectation values on a semi-classical state $|j,\what{n}\ra$ satisfying $\la \v{X}\ra= j\what{n}$, we find\footnotemark:
\be
e^{i\v{p}.\la\v{X}\ra}= e^{ijl_P\v{p}.\what{n}}
=\left(\cos (l_P\v{p}.\what{n}) +i\sin (l_P\v{p}.\what{n})\right)^j,
\ee
\be
\la  \exp(i\v{p}.\v{X})\ra =
\left( \cos \f{l_P |\v{p}|}{2} +i\f{\v{p}.\what{n}}{|\v{p}|}\sin \f{l_P |\v{p}|}{2}  \right)^{2j}.
\ee
\footnotetext{
We explicitly compute
$$
\la\exp(i\v{p}\cdot\v{X})\ra
= \la j,j| g^{-1} \exp(il_P\v{p}\cdot\v{J}) g|j,j\ra
= \la j,j| \exp(il_P(g^{-1}.\v{p})\cdot\v{J}) |j,j\ra
= \left( \cos \f{l_P |\v{p}|}{2} +i\f{(g^{-1}.\v{p})\cdot\what{n_0}}{|\v{p}|}\sin \f{l_P |\v{p}|}{2}\right)^{2j},
$$
with $(g^{-1}.\v{p})\cdot\v{J}=\v{p}.\what{n}$.
}
It is straightforward to check that in the classical limit $l_P\arr 0$, the two expressions have the same asymptotical behavior as $1+ijl_P\v{p}.\what{n}$. Noting $k\equiv l_P|\v{p}|$ and $\cos\theta\equiv\v{p}.\what{n}/|\v{p}|$, we can write:
$$
\la  \exp(i\v{p}.\v{X})\ra =
\left(\cos k +\sin^2\theta \sin^2 \f{k}{2}+ i\cos\theta\sin k \right)^j
=\left(1-\sin^2\theta \sin^2 \f{k}{2}\right)^j\,e^{2ij\vphi},
$$
with $\tan\vphi\equiv \cos\theta \tan \f{k}{2}$. This is to be compared to $\exp(i\v{p}.\la\v{X}\ra)= \exp(ijk\cos\theta)=(\cos(k\cos\theta)+i\sin(k\cos\theta))^j$. The first difference is the intensity i.e the modulus of $\la  \exp(i\v{p}.\v{X})\ra$ is not one: it presents an anisotropy centered around the direction set by the momentum $\v{p}$ and increases with its norm. The modulus being always inferior to 1, it will go to 0 as the distance $j$ grows: we will have a damping of the plane wave as the distance from the origin grows, which should be a visible effect of the non-commutative geometry. The difference in the phases $2\vphi$ and $k\cos\theta$ is rather small, but will eventually blow up (linearly) as the distance $j$ gets large.


\begin{figure}[h]
\begin{center}
\includegraphics[width=6cm,angle=270]{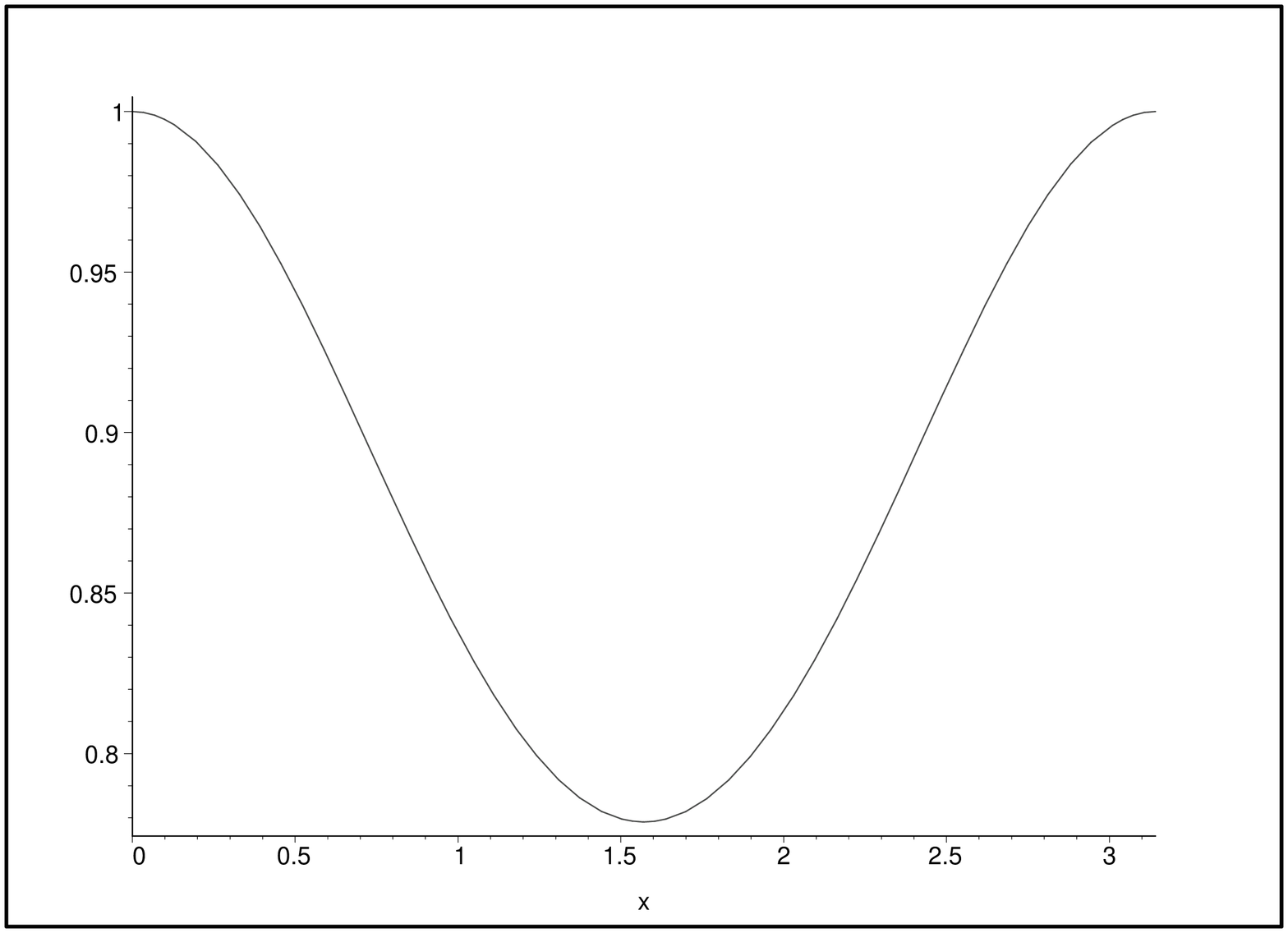}
\includegraphics[width=6cm,angle=270]{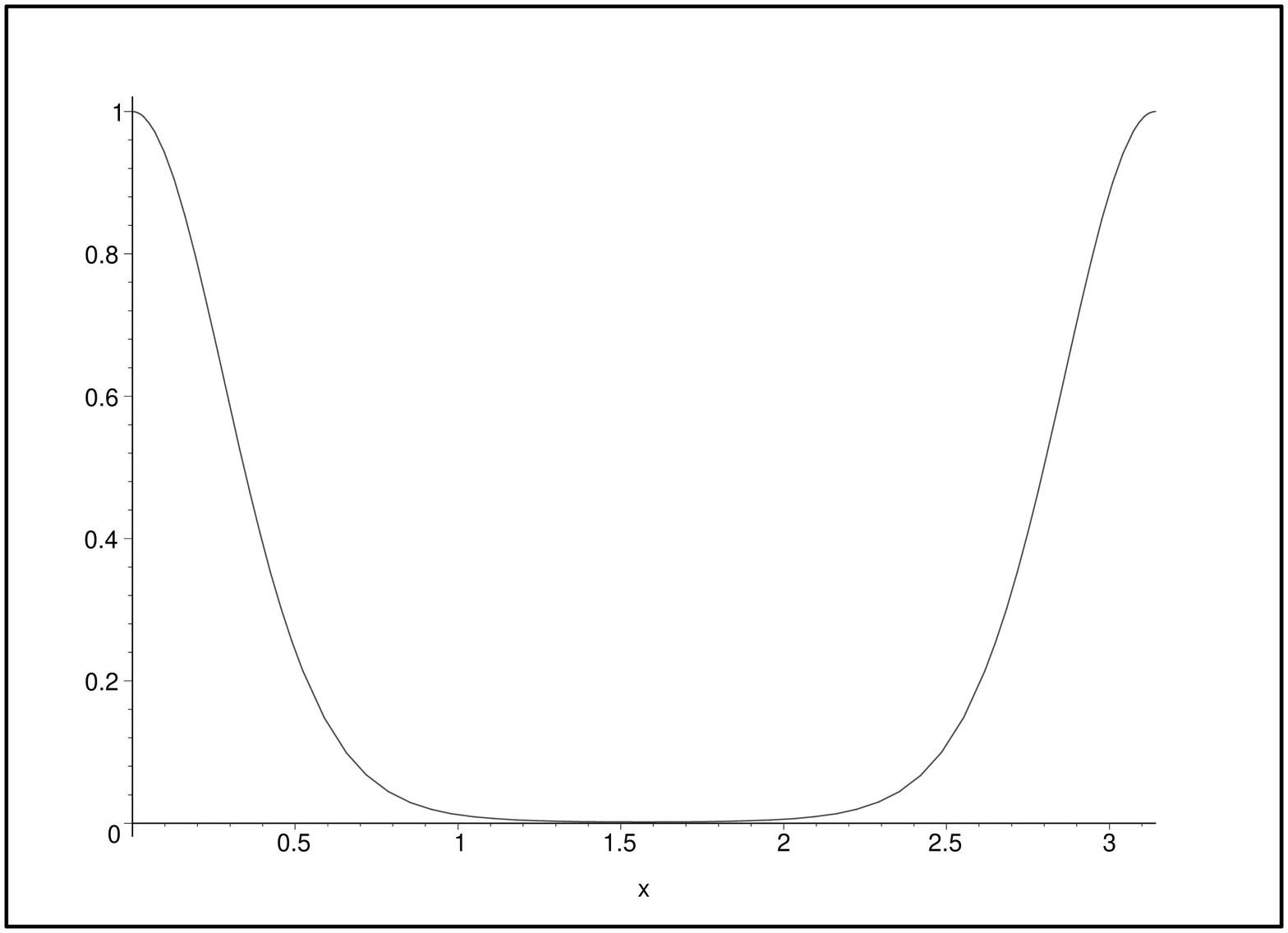}
\end{center}
\caption{Plot of $|\la\exp(i\vec{p}.\vec{X})\ra|$ for $j=100$ and $k=1/10$ then $k=1/2$
as a function of the angle $\theta$ between $\vec{p}$ and $\la\vec{X}\ra$.}
\end{figure}

\section{Coherent States: The Lorentzian 3d space-time}

Lorentzian 3d deformed special relativity is very similar to the Euclidean one. It is now the non-compact group manifold $\SU(1,1)$ that is identified as momentum space, and again one can select a preferred set of spacetime coordinate operators given by the Lie derivatives acting on this group manifold. We introduce the rotation generator $J_z$ and the two boost generators $K_x,K_y$, which satisfy the $\su(1,1)$ algebra:
$$
[J_z,K_\pm]=\pm K_\pm,\qquad
[K_+,K_-]=-2J_z,
$$
where $K_\pm=K_x\pm iK_y$. In the following, we choose to work with the $(+--)$ signature. Noting $\sigma_i$ the Lorentzian Pauli matrices representing the algebra $\su(1,1)$ in the fundamental two-dimensional representation, the group element are defined as $g=\exp(i\v{u}.\v{\sigma})$ and can always be expanded as $g=\pm\sqrt{1-\kappa^2|\v{p}|^2}+i\kappa \v{p}.\v{\sigma}$.  We distinguish two cases according to $\v{u}$ and $\v{p}$ being time-like or space-like:
\begin{itemize}
\item $\v{p}$ is time-like, $|\v{p}|^2\ge 0$, and g is a rotation of axis $\v{p}$ and angle $\theta$ such that $\sin\theta=\kappa|\v{p}|$. The sign $\pm$ can be absorbed in the definition of $\theta$.
\item $\v{p}$ is space-like, $|\v{p}|^2\le 0$, and g is a boost of axis $\v{p}$ and angle $\eta$ such that $\sinh\eta=\kappa\sqrt{-|\v{p}|^2}$.
\end{itemize}
This defines the link between the momentum $\v{p}$ living in $\mathbb{R}^{2,1}$ and the group element $g$ living in the group manifold $\SU(1,1)$. The addition of momenta is defined by the group multiplication:
$g(\v{p}_1\oplus\v{p}_2)=g(\v{p}_1)g(\v{p}_2)$ and leads to the same deformation as the Euclidean theory in the case of time-like momenta.

The spacetime coordinates are once more defined as the Lie algebra generators (see also \cite{discrete}), which produce the translations on $\SU(1,1)$:
$$
T\equiv l_P\,J_z, \quad X\equiv l_P\,K_x, \quad Y\equiv l_P\,K_y.
$$
The (spacetime) (squared) length operator is $L^2\equiv T^2-X^2-Y^2$ and is actually the Casimir operator of $\SU(1,1)$. the spectrum of $T$ is discrete $l_P\,\Z$ while the the spectra of the space coordinate remains continuous $l_P\,\R$. The spectrum of $L^2$ is more involved and consists of two distinct series:
\begin{itemize}
\item positive discrete eigenvalues: 
states $|j,m\ra$ with $j\ge 1$ correspond to the discrete principal series of irreducible representations of $\SU(1,1)$. There are two such series. The positive series have $J_z$-eigenvalues $m$ larger than $j$ while the negative series have $m\le -j$. The value of the Casimir operator is $j(j-1)$; accordingly, the eigenstates belonging to this series have the geometric interpretation of defining timelike spacetime points with respect to the origin, with the positive series corresponding to points in the future half of the lightcone, and the negative series to points in the past half of it;
\item negative continuous eigenvalues:
states $|s,m\ra$ with $s\in\R$ correspond to the continuous principal series of irreducible representations of $\SU(1,1)$. $J_z$-eigenvalues $m$ takes all values in $\Z$. The value of the Casimir operator is $-(s^2+1/4)$, and accordingly the the corresponding eigenstates are interpreted as spacelike spacetime points.
\end{itemize}
We have therefore a quantum spacetime with a discrete time-like structure; moreover, while we have a minimal space-like space-time distance $L^2=-1/4$, the space coordinates $X$ and $Y$ can take any value in $\R$. This situation is thus exactly the same as that related to the length operator in (2+1)-d Loop Quantum Gravity \cite{length3d}, as to be expected since the quantum geometric structure is based on the group $SU(1,1)$ also in that case.

Now we are interested in constructing states with minimal spread. We define the uncertainty as previously:
\be
\delta L=\sqrt{|\la L^2\ra-\la T\ra^2 +\la X\ra^2+\la Y\ra^2|}.
\ee
We consider the time-like case first. 
We can easily evaluate the distance and uncertainty on the basis states $|j,m\ra$, $m\ge j$:
$$
L_{j,m}= l_P\sqrt{j(j-1)},\qquad
(\delta L)_{j,m} = l_P\,\sqrt{|j(j-1)-m^2|}.
$$
Therefore we reach the minimal spread for the lowest weight vector $m=j$ in the positive discrete series of representations. It would be the maximal weight $m=-j$ for the negative discrete series of representations.
For time-like distances, we can conclude that:
\be
(\delta L)_{\rm min}= (\delta L)_{j,j}=l_P\sqrt{j}\sim \sqrt{l_P L},
\ee
which reproduces the result from the Euclidean theory.
For space-like vectors, the basis vectors $|s,m\ra$ give:
$$
|L_{s,m}|= l_P\sqrt{s^2+\f{1}{4}},\qquad
|(\delta L)_{s,m}| = l_P\,\sqrt{s^2+\f{1}{4}+m^2},
$$
so that $|\delta L|$ is always larger than $L$. We conclude that the eigenvector basis of $T$ is not the right basis to investigate coherent states for space-like distances, or in other words points located at a spacelike distance from the origin (where the observer is located). The interpretation of this result is not completely clear (to us at least), but it is tempting to speculate that this is due to the fact that an observer located at the origin of the reference frame (spacetime coordinate operators) we are using has no physical mean to measure events that are spacelike to her, and accordingly no identification of points outside her lightcone is possible; therefore what seems at first sight a purely mathematical accident or a formal difficulty may instead be a result of causality restrictions and a confirmation of the consistency of the whole quantum geometric scheme.

As in the Euclidean case, we can now construct the whole system of coherent states applying Perelomov procedure, starting from the semiclassical coherent states identified above; we define time-like coherent states $|j,\what{n}\in H_\pm\ra$ where $H_\pm$ is the positive (resp. negative) unit time-like hyperboloid. To get to $|j,\what{n}\ra$, we simply act on $|j,j\ra$ by an $SU(1,1)$ group element which maps the vector $\what{n_0}=(1,0,0)$, pointing to the origin of the hyperboloid, to the vector $\what{n}$. There is a $\U(1)$ ambiguity in defining this group element, corresponding to the group transformations leaving the vectors on the hyperboloid invariant, so that the relevant group transformations are the boosts only; this gives a phase ambiguity in the definition of $|j,\what{n}\ra$. The sign $\pm$ corresponds to the choice of sign of the discrete representation we use. These states have all the nice properties of the $\SU(2)$ coherent states, including the resolution of unity. 
We can write explicitly these coherent states as:
$$
|z\ra\equiv (1-|z|^2)^j\sum_{m=0}
\left(\f{\Gamma(m+2j)}{m!\Gamma(2j)}\right)^{\f{1}{2}}z^m\,|j,j+m\ra,
$$
where $\Gamma$ is the Euler $\Gamma$-function. Then we have the resolution of identity:
$$
\f{2j-1}{\pi}\int \f{{\rm d}^2z}{(1-|z|^2)^2}\, |z\ra\la z|\,=\,{\rm Id}_j, 
$$
meaning that we exhaust the set of points at timelike spacetime distance $l_P \sqrt{j(j-1)}$,
with the notation:
$$
\what{n}=(\cosh\eta,\sinh\eta\cos\phi,\sinh\eta\sin\phi),
\qquad
z=\tanh\f{\eta}{2}e^{-i\phi}.
$$
Of course we have also a full resolution of unity for all timelike distances, meaning that these system of coherent states spans geometrically the whole lightcone centered at the origin. 
In addition, one can compute the overlap between two such states:
\be
|\la j,\what{n_1}|j,\what{n_2}\ra|^2= \left(\f{1+\what{n_1}\cdot\what{n_2}}{2}\right)^{-2j}.
\ee
We see that the whole construction and its results are the exact analogue of those of the Euclidean theory, for these discrete timelike quantum points. More details can be found in \cite{cohstates}, where it is also described how to build a system coherent states for the continuous representations using functions over the Lobachevsky plane $\SU(1,1)/\U(1)$, using a different basis of states, i.e. eigenstates of different spacetime coordinate operators, in agreement with the geometric interpretation proposed above.

To sum up, the quantum spacetime of Lorentzian DSR is a non-commutative spacetime with a discrete time structure. In this context, we have reviewed the construction of coherent states for $\SU(1,1)$ and established a notion of semi-classical quantum points for the Lorentzian DSR theory for time-like separations. We obtain a spacetime distance uncertainty relation similar to the Euclidean theory with $\delta L \sim \sqrt{L}$. This is in agreement with the expectation from the holographic principle.

\section{Coherent States for the Snyder Theory: towards four dimensions}


Now, we would like to extend the 3d results to the 4d DSR theory. Also in this case we would like to define a notion of semi-classical quantum points in terms of appropriate coherent states and check whether we obtain again an uncertainty relation of the type $\delta L \sim \sqrt{L}$ similarly to the 3d theory (actually a generic feature of Lie algebra constructions) or something of the type $\delta L \sim L^{1/3}$ as expected from various arguments in 4d (including the holographic principle) or something completely different. The Snyder construction\cite{snyder}, that gives a version of DSR in 4d, is based on the definition of momentum space as the quotient space $\SO(4,1)/\SO(3,1)$, i.e. a manifold with a de Sitter geometry\footnote{One can also work with the AdS geometry given by the homogeneous space $\SO(3,2)/\SO(3,1)$.}. It is not obvious that the framework of coherent states for Lie groups can be straightforwardly applied to this case. As a start, we propose here to study using such framework the 3d Snyder theory defined from the quotient space $\SO(4)/\SO(3)$ in the Riemannian context and $\SO(3,1)/\SO(2,1)$ in the Lorentzian context, and compare the results so obtained to those that come from using instead the $SU(2)$ group structure, that have been described in the previous sections and that can be independently confirmed as trustworthty by more fundamental quantum gravity approaches, as we have discussed above.

\subsection{The Euclidean theory}
We now work with a momentum space defined as the homogeneous space $\SO(4)/\SO(3)\sim {\cal S}_3$ and we define configuration space by duality with respect to it.
More precisely, starting with the Lie algebra $\lalg{so}(4)$, we
identify the "boost" generators $K_i$ with the spacetime coordinates
$x_i/l_P$, which now truly become operators $\what{x}_i$ acting on the Hilbert space of wave functions given by square integrable functions on momentum space. The $\SO(3)$
generators $J_i$ act as usual on $\vec{x}$ which still behaves as a
vector. As previously, the coordinates are the generators of the translations on
$\SO(4)/\SO(3)$ -or equivalently basis of vectors on the tangent space- and are non-commutative:
\be
[x_i,x_j]=il_P^2\epsilon_{ijk}J_k,
\ee
although they do not form a closed Lie algebra anymore.
They have a discrete spectrum $\Z\, l_P$.

As said above, the Hilbert space of our theory is the space of $L^2$ functions over
the momentum space $\SO(4)/\SO(3)\sim {\cal S}_3$. Through harmonic
analysis, such functions can be decomposed in $\SO(4)$
representations and more precisely in {\it simple} representations of
$\SO(4)$ due to the required $\SO(3)$-invariance.
A basis of functions is given by the matrix elements:
\be
f_{n,j,m}(g\in\SO(4))=\bk{n,j,m|g|n,0},
\ee
where $n$ is the label of a simple representation (for details, see the
appendix \ref{simple}) and $j,m$ are the labels of the canonical
$\SO(3)$ basis of the considered $\SO(4)$ representation.
The vector $\ket{n,0}=\ket{n,j=m=0}$ is the $\SO(3)$-invariant vector
of the simple representation $n$ and ensures that $f_{n,j,m}$ is truly
a function on the coset $\SO(4)/\SO(3)$. We can sum up this in a
compact way, writing our Hilbert space as the direct sum of all
$\SO(4)$ simple representations $R^n$:

$$
\hh=\bigoplus_{n\in\N}R^n_{\SO(4)}, \qquad R^n_{\SO(4)}=\bigoplus_{j=0}^{2n}V^j_{\SO(3)}.
$$

\medskip

In this setting, we define the spacetime (square) length operator:
\be
\what{L}^2=x_ix_i=l_P^2\,\vec{K}^2.
\ee
Writing $\vec{K}^2=(\vec{J}^2+\vec{K}^2)-\vec{J}^2$, this operator
$\what{L}$ is obviously diagonalized in the $\ket{n,j,m}$ basis and
its spectrum is:
\be
L^2\,\ket{n,j,m} \,=\, l_P^2\,\left(4n(n+1)-j(j+1)\right)\,\ket{n,j,m}.
\ee
As $0\le j\le 2n$, the eigenvalues of $L^2$ are always positive as expected.

\subsubsection{Coherent states around the origin}

We would like to identify states localized around some classical
spacetime coordinates $(x_0,x_1,x_2)$ and to compute their spread or
uncertainty. We can start by calculating the mean value of $\what{x}_i$ on the states $\ket{n,j,m}$.  In
the following, we choose a particular value $n$ and keep it fixed.
First, one can check that (see the appendix \ref{simple} for details):
\be
\bk{j,m|\vec{x}|j,m}=l_P\,\bk{j,m|\vec{K}|j,m}=0,
\ee
which in fact results from that we are only using {\it simple}
representations. Therefore these states $\ket{j,m}$ are centered
around the origin $(0,0,0)$. It is also easy to get:
\be
\bk{j,m|J_3|j,m}=m, \quad \bk{j,m|J_{1,2}|j,m}=0.
\ee
We can define an invariant uncertainty, which is the Lie algebra
uncertainty\cite{cohstates}:
\be
\Delta=\bk{\vec{J}^2+\vec{K}^2}-\bk{\vec{J}}^2-\bk{\vec{K}}^2,
\ee
which gives:
\be
\Delta_{j,m}=4n(n+1)-m^2.
\ee
This uncertainty is minimal when for the highest weight $m=j=2n$, for
which we have $\Delta_{\rm min}=4n$.
Nevertheless, it is not exactly the uncertainty which interests
us. Physically, we are more interested into the position uncertainty:
\be
(\delta x)^2=\bk{\vec{x}^2}-\bk{\vec{x}}^2=l_P^2\,(\bk{\vec{K}^2}-\bk{\vec{K}}^2).
\ee
On the states $\ket{j,m}$, the mean value of the position vanishes so
that $(\delta x)^2=L^2$:
\be
(\delta x)^2_{j,m}=l_P^2\,(4n(n+1)-j(j+1)),
\ee
which is minimal for the largest representation $j=2n$, for which we have:
$$
(\delta x)^2_{\rm min}=2n\,l_P^2.
$$

From these results, it appears that the length spectrum is quite
misleading. All the length eigenvectors $\ket{n,j,m}$ are centered
around the spacetime origin and the value of $\what{L}^2$ gives
the (square of the) uncertainty/spread of the state.

\subsubsection{Localizing arbitrary points}
We have just seen that states in the canonical basis for simple
representations of $\SO(4)$ can be interpreted as giving a definition
of a quantum point defining the origin of our non-commutative
spacetime. 
The next task is to identify coherent states that correspond to
arbitrary (quantum) points in spacetime, localized with greater or
lesser degree of uncertainty.  
The idea is to consider arbitrary combinations of basis states, and
more precisely states that can be obtained from these by acting
with generic group elements (in the appropriate representation, of
course). In other words we will be considering states of the type:
$U^{(n)}(g)\,| n,j,m\rangle$ and consider the mean value of the
coordinate distance operators on them, in order to check that they
correspond indeed to points distinguished from the origin, and then
the corresponding dispersion in these operators to check their degree
of localization. 

In this Riemannian setting, we consider group elements $U^{(n)}(g)$ corresponding to the
exponentiation of a given Lie algebra element $K_i$. For $K_1$, we can compute\footnotemark:
$$
e^{-i \theta K_1}K_1e^{+i \theta K_1}=K_1,\quad
e^{-i \theta K_1}K_2e^{+i \theta K_1}=\ct K_2 +\st J_3,\quad
e^{-i \theta K_1}K_3e^{+i \theta K_1}=\ct K_3 -\st J_2,
$$
which allows us to derive explicitly the exact expression of the average position and spread of the state $|\theta njm \ra\equiv e^{+i \theta K_1}\,|njm\ra$:
\be
\la K_1\ra_{\theta njm} = 0, \quad
\la K_2\ra_{\theta njm} = m\st, \quad
\la K_3\ra_{\theta njm}=0, \quad
\la \vec{K} \ra^2=m^2\sin^2\theta,
\ee
\bes
(\delta K_1)^2_{\theta njm}&=&\f{1}{2}\left[
\gamma_j^2 (j^2-j+m^2) +\gamma_{j+1}^2(j^2+3j+m^2+2)
\right], \nn\\
(\delta K_2)^2_{\theta njm}&=&\f{\cos^2\theta}{2}\left[
\gamma_j^2 (j^2-j+m^2) +\gamma_{j+1}^2(j^2+3j+m^2+2)
\right],\nn\\
(\delta K_3)^2_{\theta njm}&=&\f{\sin^2\theta}{2}\left(j(j+1)-m^2\right)
+\cos^2\theta\left[
\gamma_j^2 (j^2-m^2) +\gamma_{j+1}^2((j+1)^2-m^2)
\right],\nn\\
(\delta K)^2_{\theta njm} &=&(\delta K_1)^2+\f{\sin^2\theta}{2}\left(j(j+1)-m^2\right)
+\f{\cos^2\theta}{2}\left[
\gamma_j^2 (3j^2-j-m^2) +\gamma_{j+1}^2(3j^2+7j+4-m^2)
\right],
\ees
with the factors $\gamma_j,\gamma_{j+1}$ given in appendix.
\footnotetext{To compute the action of the boosts on the Lie algebra elements, we can make use of the splitting of the $\so(4)$ algebra into left and right $\su(2)$ algebras. For example, we write:
$$
e^{-i \theta K_1}K_2e^{+i \theta K_1}=
e^{-i \theta (J^L_1 -J^R_1)} (J^L_2 - J^R_2)  e^{i\theta (J^L_1 - J^R_1)}=
\cos\theta J_2^L + \sin\theta J^L_3 - \cos\theta J^R_2+\sin\theta J^R_3=
\cos\theta K_2 +\sin\theta J_3.
$$}
Therefore, our state $|\theta n j m\ra = e^{+i \theta K_1}\,|njm\ra$ corresponds to a
point localized around the $X_2$ axis at distance $m\sin\theta$ from
the origin (remember that the labels $j$ and $m$ refer to the choice
of canonical basis with respect to the operator $J_3$).
An analogous calculation for the state $e^{i\theta K_2}\,| n j m\rangle$
would of course show that it corresponds to a
point localized around the $X_1$ axis at the same distance, and one could of course define states located at non-zero distance along the $X_3$ axis. Let us have a closer look at the uncertainty of these states. Taking as angle $\theta=\pi/2$, the expressions simplify:
$$
\la \vec{x}\ra =ml_P\,(0,1,0),\quad
(\delta x_2)=0,
$$
$$
(\delta x_1)^2=\f{l_P^2}{2}\left[
\gamma_j^2 (j^2-j+m^2) +\gamma_{j+1}^2(j^2+3j+m^2+2)
\right],
$$
$$
(\delta x_3)^2= l_P^2\,\f{1}{2}(j(j+1)-m^2).
$$
As far as $x_3$ is concerned, we have the same formulas as in the case of $\SU(2)$ coherent states. This is rather natural since $e^{+i K_1 \pi/2}$ rotates $K_2$ into $J_3$. And we reach a minimal uncertainty in $x_3$ when $m$ is set to its maximal value $j$. We then have:
\be
\la \vec{x}\ra =j\,l_P\,(0,1,0),\quad
(\delta x)=l_P\,\sqrt{\f{j}{2}+(\delta x_1)^2_{n,j,m=j}}.
\ee
We get a spread $(\delta x)$ of $(\sqrt{xl_P})/2$ outset by $(\delta x_1)$.
Then the minimal value of $\delta x_1$ is obtained for the maximal values of the labels $m=j=2n$, in which case we get:
\be
\la \vec{x}\ra =2n\,l_P\,(0,1,0),\quad
(\delta x)=l_P\,\sqrt{n+\f{n}{4}} =\f{1}{2}\sqrt{\f{5}{2}}\sqrt{xl_P},
\ee
with the same square-root law dependence of the uncertainty $\delta x$ that we had found in the 3d DSR case on the (average) distance $x$ from the origin.

A more careful analysis, allowing $\theta$ to vary and using the explicit formula for $\gamma_j$, shows that again the minimal uncertainty $(\delta x)^2$ for a fixed angle $\theta$ is again obtained for the maximal values of the labels $m=j=2n$. In this case it is straightforward to get:
$$
\la \v{x}\ra =2n\,l_P\,(0,1,0),\quad
\left(\f{\delta x}{\la x \ra}\right)^2=
\f{1}{4n}\left(1+\f{1+\cos^2\theta}{4\sin^2\theta}\right).
$$
The minimal relative uncertainty is obviously obtained when $\cos\theta=0$, i.e. when $\theta=\pi/2\,[{\rm mod}\pi]$.

Using the same techniques one can study coherent states defining arbitrary points (not located around any specific coordinate axis) in this quantum flat spacetime.
This concludes our analysis of coherent states in this Euclidean Snyder model for a non-commutative flat geometry in a 3d spacetime. We have constructed coherent states minimizing the position uncertainty around arbitrary spacetime positions and we derived a square-root law relating the (minimal) spread of localization of spacetime points to their (average) distance from the origin. At the end of the day, this seems to be a generic feature of these non-commutative 3d geometries.

\subsubsection{Relating the Snyder model to the DSR Theory}

A natural question, that the reader might ask, is the link between the earlier DSR theory using the group $\SU(2)$ as momentum space and the present Snyder theory based on the homogeneous space $\SO(4)/SO(3)\sim \ss^3$. Indeed as manifold, the two momentum spaces are isomorphic. So what is the precise difference? In fact, the Snyder formalism allows a great freedom in what one can call coordinate operators. Indeed we can choose any set of three generators $X_i$ of the translations on $\ss^3$. The standard Snyder coordinates are the boosts:
$$
X_i^{\rm Snyder}\equiv l_P\,K_i.
$$
Nevertheless, one can also define some $\su(2)$ coordinates, which we call here the DSR coordinates:
$$
X_i^{\rm DSR}\equiv l_P\,( K_i +J_i). 
$$
These are the generators of the translations on $\ss^3$ identified with the group manifold $\SU(2)$ given by the generators of the right $\SU(2)$ subgroup of the initial $\SO(4)$ symmetry group. Note that we could have also chosen the left $\su(2)$ generators $K_i-J_i$. The coherent states we have defined for 3d DSR refer indeed to this particular choice of spacetime coordinates operators.

Once the spacetime coordinate operators are chosen, we then have a freedom in the precise definition of the momentum, more exactly in the choice of mapping between points in $\ss^3$ and the standard (measurable) momentum $\vec{p}\,\in\R^3$; this is the well-known ambiguity in the choice of basis for a DSR theory (see \cite{jurek, FloEtera} for a discussion of this feature of DSR theories and for possible resolutions of this ambiguity).

\subsection{Localizing Points in 3d Lorentzian DSR spacetime}
\subsubsection{States centered around the origin}
Let us move on to the Lorentzian case and repeat the same steps as in
the previous Riemannian case.
The momentum space is now defined as the one-sheet Lorentzian
hyperboloid ${\cal H}_3\sim\SO(3,1)/\SO(2,1)$ and the position are
defined as the \lq\lq boost\rq\rq generators:
\be
x_0\equiv J_{03},\quad
x_1\equiv J_{13},\quad
x_2\equiv J_{23}.
\ee
The (squared) length $\what{L}^2\equiv J_{13}^2+J_{23}^2-J_{03}^2$ is
invariant under the 3d Lorentz group $\SO(2,1)$ and is equal to the
difference of the $\SO(3,1)$ second Casimir operator and the $\SO(2,1)$
Casimir operator. Thus its spectrum is obtained by decomposing the
$\SO(3,1)$ representations involved in the harmonic analysis over the
hyperboloid ${\cal H}_3$ into $\SO(2,1)$ representations. As explained
in appendix, such simple $\SO(3,1)$ representations are of two kinds:
we have a continuous series labeled by a real parameter $\rho\in\R_+$
and a discrete series labeled by a half-integer $n\ge 2$. Following
the decomposition of the simple representations into $\SO(2,1)$
representations given in appendix, the spectrum associated to the
continuous series is:
\be
\what{L}^2\,\ket{\rho,
  s,m}=-\left(\rho^2+1\right)+\left(s^2+\f{1}{4}\right)=s^2-\rho^2-\f{3}{4},
\ee
where $s$ runs in $\R_+$ and $m\in\Z$ is the half-integer ($U(1)$
weight) labeling the basis of the $\SO(2,1)$ representations.
The discrete simple representations give:
\be
\what{L}^2\,\ket{n,
  s,m}=\left(n^2-1\right)+\left(s^2+\f{1}{4}\right)=n^2+s^2-\f{3}{4},
\ee
with $s\in\R_+$ and $m\in\Z$, and:
\be
\what{L}^2\,\ket{n,
  j\pm,m}=\left(n^2-1\right)+\left(-j^2+\f{1}{4}\right)=n^2-j^2-\f{3}{4},
\ee
with $j\le n-1/2$ and $\pm m\ge (j+1/2)$. Let us point out the
important fact that the discrete simple representations always carry
positive $L^2$ eigenvalues which correspond to {\it spacelike}
distances. The simple continuous representations are more subtle to
interpret.

\medskip

Just as in the Euclidean case, such length eigenvector can be proved
to be centered around the origin. More precisely, one can check that:
\be
\bk{{\cal I}_3,{\cal I}_2,m|x_\mu|{\cal I}_3,{\cal I}_2,m}=0
=\bk{m|J_{01}|m}=\bk{m|J_{02}|m},\qquad
\bk{m|J_{12}|m}=m,
\ee
where ${\cal I}_d$ denotes the considered representation $\SO(d,1)$.

The uncertainty/spread\footnotemark of these states are thus entirely
exactly given by the length operator:
$(\delta x)^2_{{\cal I}_3,{\cal I}_2,m}=\bk{{\cal I}_3,{\cal
    I}_2,m|\what{L^2}|{\cal I}_3,{\cal I}_2,m}$.
Taking into account only the spacelike sector defined by the discrete
simple representations, we get that, once the $\SO(3,1)$
representation is fixed to a given ${\cal I}_3=n$, the minimal
uncertainty is obtained for the maximal value ${\cal I}_2=j=n-1/2$ for
which we have:
\be
(\delta x)_{min}=\sqrt{n^2-j^2-\f{3}{4}}=\sqrt{n-1}\ge 1.
\ee

\footnotetext{The Lie algebra "invariant" uncertainty $\Delta$ is
  computed as in the Riemannian case and is either
  $\Delta=-(\rho^2+m^2+1)<0$ or $\Delta=n^2-m^2-1$ with arbitrary
  sign.}

\subsubsection{Localize arbitrary points} 

As in the Riemannian case, quantum states to be identified with
arbitrary points in the non-commutative Lorentzian spacetime of DSR
can be obtained considering combinations of basis states defined by
applying a generic group transformation to them $U^{({\cal I}_{3})}(g)\,|{\cal I}_3,{\cal I}_{2}, m \rangle$. Again, the $\SO(2,1)$
subgroup of the $\SO(3,1)$ symmetry group of momentum space acts
trivially (in the sense that it will not change the expectation values of the coordinates $x_\mu$) and the non-trivial action is only that of transformations
generated by the operators $X_0=J_{03}$, $X_1=J_{13}$ and $X_2=J_{23}$
and their linear combinations.

Let us start by having a look at vectors of the type $\exp(+i\theta J_{13})\,|{\cal I}_3,{\cal I}_{2}, m \ra$. 
Then we can compute:
\bes
\exp(-i\theta J_{13})X_0\exp(+i\theta J_{13})&=&\ct X_0-\st J_{01}, \nn\\
\exp(-i\theta J_{13})X_1\exp(+i\theta J_{13})&=&X_1, \nn\\
\exp(-i\theta J_{13})X_2\exp(+i\theta J_{13})&=&\ct X_2+\st J_{12}.
\ees
On basis states $|{\cal I}_3,{\cal I}_{2}, m \ra$ of simple representations, only the (diagonalized) rotation $J_{12}$ have a non-zero expectation value. Therefore, similarly to the Riemannian case, we realize that a rotation $\exp(+i\theta J_{13})=\exp(+i\theta X_1)$ will generate states located around the spacetime point $(0,0,m\st)$ on the $X_2$ axis.

Let us look at the spread of these states. We will only do the explicit calculation for an angle $\theta=\pi/2$.
$(\delta X_2)$ obviously vanishes. Then we obtain:
$$
(\delta X_0)^2_{\pi/2,{\cal I}_3,{\cal I}_{2}, m}=
|\la{\cal I}_3,{\cal I}_{2}, m|J_{01}^2|{\cal I}_3,{\cal I}_{2}, m\ra|
=\f{1}{2}|{\cal Q}({\cal I}_{2})+m^2|,
$$
where ${\cal Q}({\cal I}_{2}=s)=s^2+1/4$ and ${\cal Q}({\cal I}_{2}=j\pm)=-j^2+1/4$.
Computing $(\delta X_1)$  is more complicated and requires the exact action of the $\SO(3,1)$ boost operators in the representation ${\cal I}_3$.  Focusing on $(\delta X_0)$, we see that the continuous $\SO(2,1)$ representations ${\cal I}_{2}=s$ do not lead to any interesting uncertainty result, while the discrete $\SO(2,1)$ representations ${\cal I}_{2}=j\pm$ give a minimal uncertainty when $m$ reaches its minimal value $j+1/2$. Then we obtain:
$$
\la \vec{X}\ra_{\pi/2, {\cal I}_3jm}\,=\,l_P\,\left(0,0,j+\f{1}{2}\right), \quad
(\delta X_2)=0,\quad
(\delta X_0)=l_P\,\f{1}{\sqrt{2}}\sqrt{j+\f{1}{2}}=\f{1}{\sqrt{2}}\sqrt{l_P X}.
$$
And we recover a square-root law for the spread in time of a spacetime point localized at a spacelike separation from the origin. It would be interesting to finally compute the spread $\delta X_1$ to close the analysis.

Had we rotated the basis states using $\exp(+i\theta X_2)$, we would have found a state localized around a point on the $X_1$ axis, and the analysis would have proceeded analogously. Finally, if we want to construct a state localized around a point on the timelike axis $X_0$, we would need to start with a state which is not an eigenstate of $J_{12}$. For example, we can first do a $\exp(i\eta J_{02})$ and then a $\exp(i\theta J_{13})$ (or vice-versa). Calculating the effect of such a transformation on $X_0=J_{03}$, we would get a contribution from the action of $J_{12}$, which has a non-vanishing expectation value. We leave the study of the details of this construction and of the properties of such states for future investigation.

\subsubsection{A remark on the space uncertainty}

Looking more closely at the states centered around the origin $|{\cal I}_3,{\cal I}_{2}, m \ra$, we remind that the covariant spread of a state labeled with representations of the continuous series $({\cal I}_3=\rho,{\cal I}_{2}=s)$ reads:
\be
(\delta x)^2=-l_P^2\,\left(\rho^2+1\right)+\left(s^2+\f{1}{4}\right)=l_P^2\,\left(s^2-\rho^2-\f{3}{4}\right).
\ee
This squared uncertainty on the position can be positive, or negative or even zero. Let us remind that we are in a Lorentzian metric, and that $(\delta x)^2=\delta x_\mu\delta x^\mu$ is thus positive when spacelike, negative when timelike and vanishes when null-like. We would like to insist on the fact that  $(\delta x)^2=0$ does not mean that we have perfectly localized the spacetime point and that we should look at the individual uncertainties for each coordinate to have a proper geometric interpretation of the state under consideration.

Moreover, looking at the above expression of the uncertainty $(\delta x)^2$, it is not obvious that the Planck length $l_P$ is a minimal scale for spacetime localization measurements. For this purpose, we can introduce a notion of {\it space uncertainty}:
\be
(\delta x)^2_{sp}\equiv (\delta x_1)^2 +(\delta x_2)^2.
\ee
This uncertainty is obviously not a Lorentz invariant. Nevertheless, its minimum is of course invariant.

Now the commutation relation of $x_1$ and $x_2$ is simply $[x_1,x_2]=il_P^2 J_{12}=il_P^2 J_3$; this means that $x_1/l_P$, $x_2/l_P$ and $J_{12}$ form an $\su(2)$ algebra. We can therefore perform with identical results our analysis of $\SU(2)$ coherent states: the state minimizing the uncertainty relation induced by $[x_1,x_2]=il_P^2 J_{12}$ are the coherent states $g|J, M=J\ra$, where $J$ labels the $\SU(2)$ representation, $|J,J\ra$ the maximal weight vector and $g$ an arbitrary $\SU(2)$ transformation. Computing $(\delta x)^2_{sp}$ on such states, we thus obtain:
$$
(\delta x)^2_{sp}= \,J\,l_P^2.
$$
The minimum is reached for $J=\f{1}{2}$ if working with $\SU(2)$ (or $J=1$ if working with $\SO(3)$). In the end this means that there exists a non-contractible minimal uncertainty for the localization of a point in space given by:
\be
\left[(\delta x)_{sp}\right]_{\rm min}= \,\f{1}{\sqrt{2}}l_P.
\ee
And we conclude that the Planck length $l_P$ effectively appears as a minimal length scale in space; we can not resolve the space coordinates of any event in spacetime with a better accuracy than $l_P/\sqrt{2}$.

\section{Conclusions and Outlook}


In this work we have analyzed some properties of the quantum geometry corresponding to 3-dimensional DSR (deformed or doubly special relativity) in both the Riemannian and Lorentzian setting, and in particular we tackled the issue of defining a suitable notion of \lq\lq quantum points\rq\rq in the relevant quantum spacetime, out of the appropriate definition of distance and coordinate operators. DSR models are presently understood as (candidates for) effective descriptions of quantum gravity in the weak-gravity limit, that take nevertheless into account the presence of an invariant distance or energy scale (the Planck scale) coming from the full theory. Being based on a curved momentum space and on a deformed symmetry group for spacetime, the configuration space of such models, i.e. the spacetime they describe, is naturally quantum and non-commutative. A careful analysis of such quantum flat spacetime and of its quantum geometry is needed and important for basically three reasons: 1) any proper understanding of physical processes in a DSR setting that might have phenomenological interest, and thus tell us something about the underlying fundamental quantum gravity theory, requires a clear spacetime description, as all the processes we are interested in from a physical perspective are understood as happening in spacetime; 2) the role of the Lorentz group and the way Lorentz invariance is realized, or modified to be compatible with the Planck scale, in quantum gravity is best studied in spacetime terms, since our most intuitive notion of Lorentz transformations is as spacetime symmetries; 3) the properties of the DSR quantum spacetime and of the DSR quantum geometry can be compared with those unveiled by more fundamental theories of quantum gravity and with more general features of a quantum spacetime as suggested by other approaches (such as loop quantum gravity or spin foam models or string theory); on the one hand this can suggest us which of these approaches is on the right track (by giving us means to analyze in a simplified context and then test experimentally some of their predictions) or which features of a quantum spacetime we should hope to describe with a more fundamental theory; on the other hand this can help us to understand how a DSR effective description may arise from any of these more fundamental theories in a rigorous way.    

\medskip

The second issue, regarding the role and implementation of Lorentz invariance, was studied in \cite{discrete}, together with the geometric interpretation and some properties of spacetime coordinate operators in DSR, thus starting off the analysis of its quantum spacetime geometry. In this paper we have carried this analysis further by studying the basic ingredient of a spacetime: points. The starting point was the idea that in quantum spacetime, such as the DSR one, the only notion of a \lq\lq point\rq\rq is that of a coherent state corresponding to a given non-zero values of the distance (spacetime coordinate) operators, and which is semi-classical in the sense that it minimizes the uncertainty in the simultaneous determination (measurement) of such operators. Therefore we proceeded to construct these states and to study their properties looking for a consistent geometric interpretation of them. 

\medskip

In the case of Riemannian 3d DSR the construction is based on the fact that the DSR momentum space is an $\ss^3$ manifold endowed with and $\SU(2)$ group structure, so that the corresponding Hilbert space of wave functions is $L^2(\ss^3)$ endowed with an appropriate action of $\SU(2)$; using the harmonic analysis on $\ss^3$ and the representation theory of $\SU(2)$ we have constructed the system of coherent states for this Hilbert space and identified the semiclassical ones, and we have discussed in detail the geometric meaning of their properties. The same construction was performed in the slightly more involved case of Lorentzian 3d DSR, based instead on the $\SU(1,1)$ group structure; here a careful distinction of timelike and spacelike structures and points is needed and to see how this is realized at the quantum level is very interesting; we have indeed performed again the whole construction of coherent states and studied with the same tools as in the Riemannian case their quantum geometric properties, and again uncovered the \lq\lq point structure\rq\rq of the DSR quantum spacetime, this time revealing also its quantum causality in terms of ligthcones.

\medskip

Both in the Riemannian and Lorentzian case we have pointed out an interesting square-root dependence of the quantum localization uncertainty, i.e. of the spread of coordinate operators for semiclassical coherent states (quantum points) around their mean values (classical points), on the distance itself. This is what one would have expected from arguments based {\it holographic principle} and may thus be a hint of a kind of {\it intrinsic fundamental holography} of a quantum spacetime.

\medskip

As a preliminary step towards analyzing the quantum spacetime geometry of 4d DSR, we have then considered coherent states for the Hilbert spaces given by $L^2(\SO(4)/\SO(3))$ and $L^2(\SO(3,1)/\SO(2,1))$, where the relevant group actions are represented by $\SO(4)$ and $\SO(3,1)$ respectively; this is because in 4d DSR momentum space is indeed given by 4d De Sitter space (or the 4-sphere in the Riemannian case) with the De Sitter (or $\SO(5)$) group structure. In this context we have again constructed the system of coherent states and analyzed their properties and their geometric interpretation, comparing the results with those obtained previously; we have shown how the construction proceeds in the so-called Snyder coordinates (the construction of coherent states in $\kappa$-Minkowski coordinates was performed in the 4-dimensional case in \cite{discrete}). The extension of this construction and these results to the 4-dimensional case is more technically and computationally involved but would proceed in complete analogy.  

\medskip

Our analysis and results nicely complement, we think, previous work on the quantum geometry of non-commutative spacetimes, and on the notion of points in such a context. In particular we have in mind the important work of Majid \cite{majidCoh} for the fuzzy sphere and that of Toller \cite{toller} for the Snyder spacetime. In fact, we have constructed the same coherent states and differential calculus that Majid studied for the $\su(2)$ Lie algebra treated as a non-commutative space, but using different tools, ours being simply based on the representation theory of Lie groups and not on the theory of quantum groups; also we have seen that our identification of quantum points (coherent states) uses the tool of POVMs, which is the main tool used also by Toller in his analysis. It would be very interesting to deepen the comparison between these various approaches with ours and to extend in this way our results.

\appendix

\section{Simple representations of $\SO(4)$}
\label{simple}

$\lalg{so}(4)\sim\lalg{spin}(4)$ is isomorphic to two copies of $\lalg{su}(2)$, $\lalg{su}(2)_L\oplus\lalg{su}(2)_R$, as a Lie algebra. Its representation are labeled by two spin $(j_L,j_R)$. If we call $\vec{J}_L$ and $\vec{J}_R$ the standard generators of the left and right $\SU(2)$ groups, the generators of the space rotation group $\SU(2)$ are $\vec{J}=\vec{J}_L+\vec{J}_R$ while the "boosts" generators are $\vec{K}=\vec{J}_L-\vec{J}_R$.
The two Casimir operators are:
$$
C_1=\vec{J}^2+\vec{K}^2=2\vec{J}_L^2+2\vec{J}_R^2=2j_L(j_L+1)+2j_R(j_R+1),
$$
and
$$
C_2=\vec{J}\cdot\vec{K}=\vec{J}_L^2-\vec{J}_R^2=j_L(j_L+1)-j_R(j_R+1).
$$
The so-called {\it simple representations} are equivalently defined as the irreducible representations which contain a $\SU(2)$-invariant vector or such that $C_2=\vec{J}\cdot\vec{K}=0$. Therefore, they are such that $j_L=j_R$ and we label them by a single half-integer $n=j_L=j_R$.
For a given $n$, the action of the generators in the standard $\SU(2)$ basis is:
\bes
J_3\ket{j,m}&=&m\,\ket{j,m} \nn\\
J_+\ket{j,m}&=&\sqrt{(j-m)(j+m+1)}\,\ket{j,m+1}\nn\\
J_-\ket{j,m}&=&\sqrt{(j+m)(j-m+1)}\,\ket{j,m-1}\nn\\
K_3\ket{j,m}&=&\gamma_j\sqrt{j^2-m^2}\,\ket{j-1,m}+\gamma_{j+1}\sqrt{(j+1)^2-m^2}\,\ket{j+1,m} \nn\\
K_+\ket{j,m}&=&\gamma_j\sqrt{(j-m)(j-m-1)}\,\ket{j-1,m+1}+\gamma_{j+1}\sqrt{(j+m+1)(j+m+2)}\,\ket{j+1,m+1} \nn\\
K_-\ket{j,m}&=&\gamma_j\sqrt{(j+m)(j+m-1)}\,\ket{j-1,m-1}+\gamma_{j+1}\sqrt{(j-m+1)(j-m+2)}\,\ket{j-1,m+1},
\ees
with
$$
\gamma_j=\f{1}{2}\sqrt{\f{n(n+1)-\f{1}{4}(j-1)(j+1)}{\left(j-\f{1}{2}\right)\left(j+\f{1}{2}\right)}}.
$$
As $\gamma_{j=2n+1}=0$, the $\SU(2)$ spin $j$ runs from 0 to $2n$. Then it is straightforward to check that
$C_1=J^2+K^2=4n(n+1)$.

\section{Simple representations of the Lorentz group $\SO(3,1)$}
\label{simpleL}

An irreducible representation of the Lorentz group is characterized
by two numbers $(n,\mu)$, where $n\in \N/2$ and $\mu\in \C$.
The unitary representations correspond to two cases:
\begin{enumerate}
\item the principal series:
$(n,\mu)=(n,i\rho),  \quad n\in \N/2, \ \rho\in \R$.
\item the supplementary series:
$(n,\mu)=(0,\rho), \quad |\rho|<1, \ \rho\in \R$.
\end{enumerate}
The principal series representations are the ones intervening
in the Plancherel decomposition formula for $L^2$ functions over $\SL(2,\C)$.
The Casimir are then:
$$
C_1=\vec{K}^2-\vec{J}^2=\,\rho^2-n^2+1, \qquad
C_2=\vec{J}\cdot\vec{K}=\, 2n\rho.
$$
Simple representations are defined as the representations of the principal series
with the vanishing Casimir  $C_2(\lalg{sl}(2,C))$.
There are clearly two types of such representations: a discrete series $(n,0)$ and a continuous series $(0,i\rho)$.

Let us decompose these simple representations on $\SU(2)$ representations. In general, a $(n,i\rho)$ representations will decompose onto spins $j\ge n$. Then we have:
\bes
J_3\ket{j,m}&=&m\,\ket{j,m} \nn\\
J_+\ket{j,m}&=&\sqrt{(j-m)(j+m+1)}\,\ket{j,m+1}\nn\\
J_-\ket{j,m}&=&\sqrt{(j+m)(j-m+1)}\,\ket{j,m-1}\nn\\
K_3\ket{j,m}&=&\gamma_j\sqrt{j^2-m^2}\,\ket{j-1,m}-\gamma_{j+1}\sqrt{(j+1)^2-m^2}\,\ket{j+1,m} \nn\\
K_+\ket{j,m}&=&\gamma_j\sqrt{(j-m)(j-m-1)}\,\ket{j-1,m+1}+\gamma_{j+1}\sqrt{(j+m+1)(j+m+2)}\,\ket{j+1,m+1} \nn\\
K_-\ket{j,m}&=&-\gamma_j\sqrt{(j+m)(j+m-1)}\,\ket{j-1,m-1}-\gamma_{j+1}\sqrt{(j-m+1)(j-m+2)}\,\ket{j-1,m+1},
\ees
with
$$
\gamma_{(j)}=\frac{i}{2}
\sqrt{\frac{j^2-n^2}{\left(j-\f{1}{2}\right)\left(j+\f{1}{2}\right)}}
\qquad\textrm{or}\qquad
\gamma_{(j)}=\frac{i}{2}
\sqrt{\frac{j^2+\rho^2}{\left(j-\f{1}{2}\right)\left(j+\f{1}{2}\right)}}.
$$

Irreducible $\SL(2,\C)$ representations containing a vector invariant under the compact subgroup $\SU(2)$ or under the non-compact subgroup $\SU(1,1)$ are very naturally simple representations. Looking at the harmonic analysis over the hyperboloid $\SL(2,\C)/\SU(2)\sim \SO(3,1)/\SO(3)$, $L^2$ functions decompose onto simple representation from the continuous series $(0,i\rho)$. On the other hand, looking at the harmonic analysis over the hyperboloid $\SL(2,\C)/\SU(1,1)\sim\SO(3,1)/\SO(2,1)$, we need all the simple representations both from the continuous and discrete series. This latter case is the one relevant for the study of the Lorentzian 3d DSR theory.

\medskip

Let them look at the decomposition of simple representations into $\SO(2,1)$ representations. To start with, $\SO(2,1)$ irreducible unitary representations are of two kinds (at least dealing wit the principal series):
\begin{enumerate}
\item the (principal) continuous series labeled by a real parameter $s\in\R_+$:
the Casimir $C(\lalg{so}(2,1))=J_{01}^2+J_{02}^2-J_{12}^2$ is simply $s^2+1/4$, and the representations have $U(1)$ weights $m\in\Z$ .
\item the (principal) positive and negative series series labeled by an integer parameter $j\in\N, j\ge 1$ (and a sign $\pm$) :
the Casimir is then $C(\lalg{so}(2,1))=-j^2+1/4$, positive representations have $U(1)$ weights $m$ running from $j+1/2$ to $+\infty$ while negative representations have weights running from $-(j+1/2)$ to $-\infty$.
\end{enumerate}

Simple $\SO(3,1)$ representations of the continuous series decompose only onto the continuous series of $\SO(2,1)$ representations:
\be
R^{\rho}=\,2\,\int_0^\infty ds\,V^{s},
\ee
while simple representations from the discrete series decompose over all $\SO(2,1)$ representations:
\be
R^{n}=\,\bigoplus_{j=1}^{n-1/2}(V^{j+}\oplus V^{j-})\,\oplus\,\int_0^\infty ds\,V^{s}.
\ee


\end{document}